\documentclass{article}
\usepackage{epsfig}
\usepackage{color}
\usepackage{amsmath}
\usepackage{amssymb}
\usepackage{lscape}

\makeatletter
\@addtoreset{equation}{section}

\makeatother

\begin{document}

\fontsize{11pt}{0.5cm}\selectfont

\begin{flushright}
April 2019

KEK-TH-2126
\end{flushright}

\begin{center}

\vspace{3cm}

{\LARGE 
\begin{center}
Linear Chern-Simons-matter Theories

\vspace{5mm}

in the Planar Limit
\end{center}
}

\vspace{2cm}

Takao Suyama \footnote{e-mail address: tsuyama@post.kek.jp}

\vspace{1cm}

{\it 
KEK Theory Center, High Energy Accelerator Research Organization (KEK), 

Oho 1-1, Tsukuba, Ibaraki 305-0801, Japan
}

\vspace{2cm}

{\bf Abstract} 

\end{center}

We study ${\cal N}=3$ linear Chern-Simons-matter theories in the planar limit. 
The matter content of the theory is depicted by a linear-shape diagram with $n$ nodes and $n-1$ links for any $n$. 
The free energy and the vevs of BPS Wilson loops are given in terms of a single 1-form on $\mathbb{CP}^1$ which can be determined explicitly for all linear theories. 
The analytic structure of the vevs of the Wilson loops is investigated in detail for $n=1$ and $n=2$. 
The addition of fundamental matters is also discussed. 

\newpage
\vspace{1cm}

\section{Introduction}

\vspace{5mm}

A Chern-Simons-matter theory appears as the worldvolume theory on a D-brane system. 
For example, one may construct a three-dimensional gauge theory, without Chern-Simons terms, by a brane construction using D3-branes intersecting with NS5-branes \cite{Hanany:1996ie}. 
The gauge group is of the form $G=\prod_{a=1}^n{\rm U}(N_a)$. 
Chern-Simons terms are induced by replacing the $i$-th NS5-brane in the D3-NS5 system to a $(1,\tilde{k}_i)$5-brane \cite{Kitao:1998mf}\cite{Bergman:1999na}. 
Some of the integers $\tilde{k}_i$ can be zero. 
The Chern-Simons level $k_a$ for each ${\rm U}(N_a)$ gauge factor is determined by these $\tilde{k}_i$. 

\begin{figure}
\begin{center}
\includegraphics{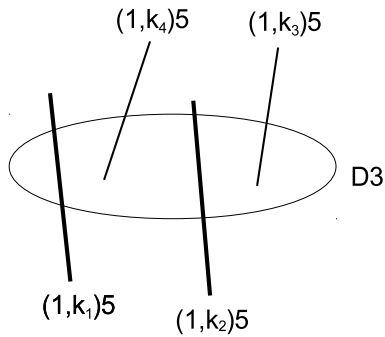} \label{brane}
\end{center}
\caption{The brane configuration which gives a Chern-Simons-matter theory with $n=4$. }
\end{figure}

Let us consider this brane system in $\mathbb{R}^{1,8}\times S^1$ in which the D3-branes wrap on $S^1$. 
This is depicted in Figure \ref{brane}. 
In the worldvolume theory, the number $n$ of the ${\rm U}(N_a)$ factors in $G$ corresponds to the number of the  $(1,\tilde{k}_i)$5-branes, and $N_a$ are the numbers of D3-brane segments suspended between successive $(1,\tilde{k}_i)$5-branes. 
Via a chain of dualities, this system can be related to multiple M2-branes in some background. 
This relation enables us to determine the worldvolume theory on the M2-branes. 
ABJM theory \cite{Aharony:2008ug} corresponds to the case $n=2$. 
The theories with $n>2$ were also discussed in \cite{Imamura:2008nn}\cite{Jafferis:2008qz}. 
Their gravity duals are M-theory on backgrounds of the form ${\rm AdS}_4\times M_7$ where $M_7$ is a  seven-dimensional 3-Sasakian manifold. 

\vspace{5mm}

In this paper, we consider a family of Chern-Simons-matter theories with ${\cal N}=3$ supersymmetry, each of which consists of matter fields belonging to the bi-fundamental representation of ${\rm U}(N_a)\times{\rm U}(N_{a+1})\subset G$. 
This family includes the ${\cal N}=4$ theories of \cite{Gaiotto:2008sd}\cite{Hosomichi:2008jd}. 
The theory is usually depicted by a diagram such as Figure \ref{diagram}. 
In this paper, we refer to a theory of this kind as a linear theory, due to the shape of the diagram. 
More details on this theory is given in section \ref{preliminaries}. 

One may regard the linear theory as a deformation of ABJM theory or its generalization mentioned above. 
Namely, one may take a limit $N_a\to0$ for which a set of D3-brane segments disappears, or another limit $k_a\to\infty$ for which one ${\rm U}(N_a)$ factor in $G$ becomes a global symmetry since $1/k_a$ is the coupling constant of the gauge field for the ${\rm U}(N_a)$ factor. 
In the latter case, the resulting theory also has matter fields belonging to the fundamental representation of ${\rm U}(N_{a-1})$ and ${\rm U}(N_{a+1})$ factors in $G$. 
It would be natural to expect that the gravity dual of a linear theory could be obtained by taking a suitable limit in the gravity dual \cite{Aharony:2008ug}\cite{Imamura:2008nn}\cite{Jafferis:2008qz} of ABJM theory or its generalization. 
However, the limit in the gravity dual seems to involve a large flux or a singular geometry, and it is not clear whether such a dual theory is useful for further analysis. 
Interestingly, as we will show in this paper, the field theory analysis of the linear theories turns out to be quite simple and explicit, even in the strong coupling region. 
Therefore, it might be possible to use the strongly coupled field theory to study a possibly singular gravity theory. 

We investigate the linear theories based on the localization formula for the partition function \cite{Kapustin:2009kz}. 
We take a limit, called the planar limit, in which the ranks $N_a$ of the ${\rm U}(N_a)$ factors and the Chern-Simons levels $k_a$ become large, keeping their ratios finite. 
This limit enables us to analyze some observables of the theories, the free energy and the vevs of BPS Wilson loops, for arbitrary values of the 't~Hooft couplings for all the linear theories. 
We obtain the formulas (\ref{formula t Hooft})(\ref{formula Wilson}) for the vevs of BPS Wilson loops as functions of the 't~Hooft couplings. 
These are given as parametric representations using simple functions, whose analytic structure can be studied easily. 
We also obtain an integral formula for the free energy of the theories. 
Their detailed analysis will be postponed to a future publication. 

\vspace{5mm}

This paper is organized as follows. 
In section \ref{preliminaries}, we describe the ${\cal N}=3$ linear Chern-Simons-matter theories in more detail. 
We also recall the localization formulas for the observables studied in this paper. 
Then, we define the planar limit for the theories. 
In section \ref{planar solution}, we show that all the information necessary in this paper is encoded in a 1-form $\Lambda$ on $\mathbb{CP}^1$, which we call the fundamental 1-form of the theory. 
The explicit form of $\Lambda$ gives simple formulas for the vevs of BPS Wilson loops. 
The integral formula for the free energy is also given in terms of $\Lambda$. 
%
The results obtained in section \ref{planar solution} are examined in detail in section \ref{examples}. 
We show that our formulas reproduce known results for $n=1$ theory, that is, ${\cal N}=3$ pure Chern-Simons theory. 
For $n=2$ theory, which was studied recently \cite{Nosaka:2017ohr}\cite{Nosaka:2018eip}, our formulas give new results. 
In section \ref{flavors}, we briefly explain how to introduce fundamental matters to the theories discussed so far. 
Section \ref{discuss} is devoted to discussion. 
Some technical details are relegated to appendices. 

\vspace{1cm}

\section{${\cal N}=3$ linear theories in the planar limit} \label{preliminaries}

\vspace{5mm}

We consider a family of Chern-Simons theories coupled to matters in three dimensions. 
We require that the theories have ${\cal N}=3$ supersymmetry, which is the largest supersymmetry allowed for arbitrary gauge groups and their representations of the matter fields. 
In such theories, the matter fields form ${\cal N}=4$ hypermultiplets. 
The general structure of the Lagrangian of those theories can be found, for example, in \cite{Gaiotto:2007qi}. 

\begin{figure}
\begin{center}
\includegraphics{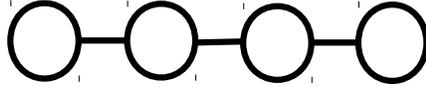} \label{diagram}
\end{center}
\caption{The diagram depicting the matter contents of the linear theory with $n=4$. }
\end{figure}

In this paper, we mainly focus our attention on theories of the following type: 
\begin{itemize}
\item The gauge group is of the form $\prod_{a=1}^n{\rm U}(N_a)_{k_a}$ where $k_a\in\mathbb{Z}$ are the Chern-Simons levels. 
\item For each factor ${\rm U}(N_a)\times{\rm U}(N_{a+1})$ ($a=1,2,\cdots,n-1$) of the gauge group, there is a single ${\cal N}=4$ hypermultiplet belonging to the bi-fundamental representation. 
\end{itemize}
These data can be depicted in Figure \ref{diagram}. 
In section \ref{flavors}, we will also briefly discuss similar theories with additional ${\cal N}=4$ hypermultiplets belonging to the fundamental representation of ${\rm U}(N_a)$. 
Note that an ${\cal N}=4$ hypermultiplet consists of two ${\cal N}=2$ chiral multiplets belonging to a  representation and its complex conjugate. 
Therefore, each link in Figure \ref{diagram} may be replaced with two arrows with opposite directions. 

We call these theories (both with and without flavors) the linear theories because of the shape of the corresponding diagram. 
The linear theory with $n=2$ were recently studied in \cite{Nosaka:2017ohr}\cite{Nosaka:2018eip}. 
In our investigation, the value of $n$ can be arbitrary. 
We will take the parameters $N_a,k_a$ to be large, while their ratios $N_a/k_b$ can be arbitrary finite values. 

As observables of interest, we consider the free energy of the theory, and the vevs of BPS Wilson loops \cite{Gaiotto:2007qi} defined as 
\begin{equation}
W_a(C)\ :=\ \frac1{N_a}{\rm Tr}\,P\exp\left[ \int_C d\tau\left( iA^{(a)}_\mu\dot{x}^\mu+\sigma^{(a)}|\dot{x}| \right) \right]
   \label{Wilson loop QFT}
\end{equation}
for each ${\rm U}(N_a)$, where $A_\mu^{(a)}$ and $\sigma^{(a)}$ are the gauge field and the real scalar field, respectively, in the ${\cal N}=2$ vector multiplet for ${\rm U}(N_a)$. 
For a suitable choice of the contour $C$, they preserve two supercharges. 

\vspace{5mm}

\subsection{Localization formulas}

\vspace{5mm}

Due to the supersymmetric localization \cite{Kapustin:2009kz}, the free energy $F$ and the vevs $W_a$ of the Wilson loops (\ref{Wilson loop QFT}) can be given in terms of finite-dimensional integrals. 

Let $n_{ab}$ be the number of bi-fundamental hypermultiplets for a factor ${\rm U}(N_a)\times{\rm U}(N_b)$ of the gauge group. 
We define $n_{aa}=0$. 
The localization formula for the partition function of an ${\cal N}=3$ Chern-Simons matter theory whose matters consists of bi-fundamental fields is given as 
\begin{equation}
Z\ =\ \int du\,\exp\left[ {\frac{i}{4\pi}\sum_{a=1}^nk_a\sum_{i=1}^{N_a}(u^a_i)^2} \right]\frac{{\prod_{a=1}^n\prod_{i<j}^{N_a}\sinh^2\frac{u^a_i-u^a_j}{2}}}{{\prod_{a,b=1}^{n}\prod_{i=1}^{N_a}\prod_{j=1}^{N_{b}}\left( \cosh\frac{u^a_i-u^{b}_j}2 \right)^{\frac12n_{ab}}}}, 
   \label{localized partition function}
\end{equation}
where $du:=\prod_{a,i}du^a_i$ is the integration measure. 
We have omitted the overall numerical factor which are irrelevant below. 
It is convenient to rewrite the integrand in (\ref{localized partition function}) as $\exp\left( -S_{\rm eff}[u] \right)$ where 
\begin{eqnarray}
S_{\rm eff}[u] 
&=& \frac{1}{4\pi i}\sum_{a=1}^nk_a\sum_{i=1}^{N_a}(u^a_i)^2-\sum_{a=1}^n\sum_{i<j}^{N_a}\log\sinh^2\frac{u^a_i-u^a_j}{2} \nonumber \\
& & +\frac12\sum_{a,b=1}^nn_{ab}\sum_{i=1}^{N_a}\sum_{j=1}^{N_b}\log\cosh\frac{u^a_i-u^b_j}2. 
\end{eqnarray}
For a linear theory, $n_{ab}$ are given as 
\begin{equation}
n_{ab}\ =\ \delta_{a,b+1}+\delta_{a,b-1}. 
\end{equation}
Note that the indices are {\it not} considered modulo $n$. 
The free energy $F$ is then given as 
\begin{equation}
F\ :=\ -\log Z. 
\end{equation}
The vevs $W_a$ of the Wilson loops (\ref{Wilson loop QFT}) are given as 
\begin{equation}
W_a\ :=\ \frac1Z\int du\,e^{-S_{\rm eff}[u]}\cdot\frac1{N_a}\sum_{i=1}^{N_a}e^{u^a_i}. 
   \label{localized Wilson loop}
\end{equation}

\vspace{5mm}

\subsection{Planar limit} \label{planar limit}

\vspace{5mm}

The integrals (\ref{localized partition function}) and (\ref{localized Wilson loop}) are too complicated to analyze exactly for generic values of the parameters. 
To proceed further, we take a limit of the parameters described below. 

We observe that $S_{\rm eff}[u]$ scales as $k^2$ for large $k$ when 
the parameters $k_a$ and $N_a$ scale as 
\begin{equation}
k_a\ =\ \kappa_ak, \hspace{1cm} 2\pi iN_a\ =\ t_ak, 
   \label{planar parameters}
\end{equation}
where $\kappa_a$ and $t_a$ are fixed constants. 
Therefore, the saddle point approximation for the integrals (\ref{localized partition function}) and (\ref{localized Wilson loop}) becomes exact in the large $k$ limit. 
We call this the planar limit. 

In the following, we call $t_a$ the 't~Hooft couplings. 
The physical values of $\kappa_a$ and $t_a$ are real and pure imaginary, respectively. 
However, in the following sections, we will regard them as complex parameters. 
Since the overall scale of $(\kappa_a,t_a)$ can be absorbed into $k$, the parameter space becomes $\mathbb{CP}^{2n-1}$. 
The physical parameter region is a $2n-1$ real-dimensional subspace. 

In the planar limit, the observables $F$ and $W_a$ are determined in terms of the solution of the saddle point equations 
\begin{equation}
\frac{\partial S_{\rm eff}}{\partial u^a_i}\ =\ \frac{k_a}{2\pi i}u^a_i-\sum_{j\ne i}^{N_a}\coth\frac{u^a_i-u^a_j}2+\sum_{b\ne a}^n\frac{n_{ab}}2\sum_{j=1}^{N_b}\tanh\frac{u^a_i-u^b_j}2\ =\ 0. 
   \label{saddle point original}
\end{equation}
Since these equations have complex coefficients, the solution $\{\bar{u}^a_i\}$ consists of complex values in general. 
It is known that there are infinite number of saddle points for equations of this kind \cite{Morita:2017oev}. 
In the following, we select one of the saddle points which reproduces correct results in the weak 't~Hooft coupling limit. 

\vspace{5mm}

In the context of the matrix model, it is well-known that the complex analysis is a very powerful tool for dealing with the information of the saddle point given by (\ref{saddle point original}). 
We introduce the resolvents 
\begin{equation}
v_a(z)\ :=\ \frac{t_a}{N_a}\sum_{i=1}^{N_a}\frac{z+z^a_i}{z-z^a_i}, \hspace{1cm} z_i^a\ :=\ -(-1)^a\exp(\bar{u}^a_i), 
   \label{resolvents}
\end{equation}
in terms of the solution $\{\bar{u}^a_i\}$ of (\ref{saddle point original}). 
The resolvents encode the information of the saddle point as poles at $z=z^a_i$. 

In the planar limit, the poles $z=z^a_i$ for each $a$ are expected to be accumulated, forming a segment $I_a$ of a curve in $\mathbb{C}$. 
Then, $v_a(z)$ has a square-root branch cut on $I_a$. 
The segments $I_a$ are not necessarily lying on the real axis of $\mathbb{C}$ since $\bar{u}^a_i$ are complex in general. 

The saddle point equations (\ref{saddle point original}) can be rewritten in terms of $v_a(z)$. 
In fact, it was observed in \cite{Suyama:2016nap} that their derivatives $v_a'(z)$ are more suitable quantities than $v_a(z)$ for solving the equations (\ref{saddle point original}). 
The derivatives $v_a'(z)$ satisfy 
\begin{equation}
2\kappa_a\ =\ xv_a'(x_+)+xv_a'(x_-)-xv_{a-1}'(x)-xv_{a+1}'(x), \hspace{1cm} x\in I_a, 
   \label{saddle point derivative}
\end{equation}
where we defined $v_0(z)=v_{n+1}(z)=0$, and $x_\pm$ are points in the vicinity of $x$ above or below $I_a$. 
When $I_a$ is a segment on the real axis, then $x_\pm=x\pm i0$. 
A derivation of these equations can be found in Appendix \ref{resolvent equations}. 

\vspace{5mm}

Once $v_a'(z)$ are obtained, we can calculate the observables $F$ and $W_a$. 
The vevs $W_a$ are given rather simply as 
\begin{equation}
W_a\ =\ \frac{(-1)^a}{2t_a}\lim_{z\to\infty}z\cdot zv_a'(z). 
\end{equation}
The free energy $F$ can be obtained as follows. 
Recalling that $S_{\rm eff}[u]$ scales as $k^2$ in the planar limit, we notice that the free energy $F$ is a homogeneous function of $k_a$ and $N_a$ of degree two. 
Therefore, $F$ satisfies 
\begin{equation}
F\ =\ \frac12\sum_{a=1}^n\left( k_a\frac{\partial F}{\partial k_a}+N_a\frac{\partial F}{\partial N_a} \right). 
\end{equation}
Note that the finite quantity in the planar limit is $F/k^2$ which can be written in terms of $\kappa_a$ and $t_a$. 
The derivatives of $F$ can be obtained from $v_a'(z)$ as 
\begin{eqnarray}
\frac{\partial F}{\partial k_a} 
&=& \frac{iN_a}{24\pi t_a}\int_{C_a}\frac{dz}{2\pi i}\frac{zv_a'(z)}{z}(\log (\epsilon_az))^3, \\ [2mm]
\frac{\partial F}{\partial N_a} 
&=& \int_{p_a}^\infty dz\left[ -\frac{N_a}{t_a}\frac{zv_a'(z)}{z}\log (\epsilon_az)+\sum_{b\ne a}^n\frac{n_{ab}}{2}\frac{N_b}{t_b}\frac{zv_b'(z)}{z}\log (\epsilon_az) \right], 
\end{eqnarray}
where $C_a$ is a closed contour encircling $I_a$, and $p_a$ is an endpoint of $I_a$. 
The details of these formulas can be found in Appendix \ref{observables}. 

Note that there is no dependence of the equations (\ref{saddle point derivative}) on the 't~Hooft couplings $t_a$. 
As will be shown in the next section, the equations (\ref{saddle point derivative}) for fixed $\kappa_a$ determine $v_a'(z)$ up to $n$ complex parameters, say $\xi_a$. 
Then, the observables $F$ and $W_a$ are given as functions of $\kappa_a$ and $\xi_a$. 
They can be converted to functions of $\kappa_a$ and $t_a$ by using 
\begin{equation}
v_a(\infty)\ =\ -v_a(0)\ =\ t_a
\end{equation}
which relate $\xi_a$ to $t_a$. 
In terms of $v_a'(z)$, we can use instead 
\begin{equation}
t_a\ =\ \frac12\int_0^\infty \frac{dz}z\cdot zv_a'(z). 
   \label{t Hooft}
\end{equation}

In the next section, we will find that $t_a$ and $W_a$ are simple functions of $\kappa_a$ and $\xi_a$. 
On the other hand, $F$ is given by an integral of elementary functions. 
We will mainly discuss the properties of $W_a$ in section \ref{examples}, while the investigation of $F$ will appear in a future publication. 

\vspace{1cm}

\section{Planar solution}   \label{planar solution}

\vspace{5mm}

In this section, we solve the equations 
\begin{equation}
2\kappa_a\ =\ xv_a'(x_+)+xv_a'(x_-)-xv_{a-1}'(x)-xv_{a+1}'(x), \hspace{1cm} x\in I_a. 
   \label{saddle point derivative 2}
\end{equation}
Recall that we set $v_0(z)=v_{n+1}(z)=0$. 

Our first step is to eliminate the constants in the left-hand side. 
We define $\omega_a(z)$ to be constant shifts of $zv_a'(z)$, that is, $\omega_a(z)$ satisfy 
\begin{equation}
zv_a'(z)\ =\ c_a+\omega_a(z). 
\end{equation}
We choose $c_a$ to be a solution of 
\begin{equation}
2\kappa_a\ =\ 2c_a-c_{a-1}-c_{a+1}, 
   \label{def of c_a}
\end{equation}
where we set $c_0=c_{n+1}=0$. 
These equations always have a solution. 
It is interesting to notice that these equations can be written as 
\begin{equation}
(2\kappa_1,\cdots,2\kappa_n)\ =\ (c_1,\cdots,c_n)A, 
\end{equation}
where $A$ is the Cartan matrix of $su(n+1)$ which is non-degenerate \cite{Suyama:2017rfh}. 
For this choice of $c_a$, we find that $\omega_a(z)$ satisfy 
\begin{equation}
\omega_a(x_+)+\omega_a(x_-)-\omega_{a-1}(x)-\omega_{a+1}(x)\ =\ 0, \hspace{1cm} x\in I_a,
   \label{homogeneous}
\end{equation}
where $\omega_0(z)=\omega_{n+1}(z)=0$. 

The equations (\ref{homogeneous}) can be further simplified by introducing $\Omega_\alpha(z)$ ($\alpha=0,1,\cdots,n$) defined as 
\begin{equation}
\Omega_\alpha(z)\ :=\ \omega_{\alpha+1}(z)-\omega_\alpha(z). 
\end{equation}
We find that (\ref{homogeneous}) can be written as 
\begin{equation}
\Omega_a(x_\pm)\ =\ \Omega_{a-1}(x_\mp), \hspace{1cm} x\in I_a, 
   \label{gluing}
\end{equation}
for $a=1,2,\cdots,n$. 

\vspace{5mm}

\subsection{A function on $\mathbb{CP}^1$} \label{pull-back}

\vspace{5mm}

The following geometric consideration is useful for solving the equations (\ref{gluing}). 

\begin{figure}
\begin{center}
\includegraphics{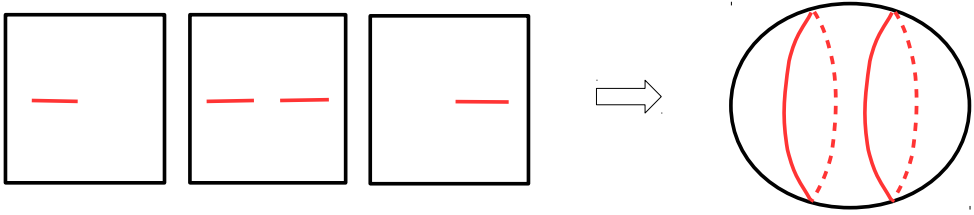}
\end{center}
\caption{The gluing of three complex planes. The branch cuts are represented by red lines. }
   \label{gluing to CP1}
\end{figure}

Let $\Sigma_\alpha:=\mathbb{C}\backslash(I_\alpha\cup I_{\alpha+1})$ be a domain of $\mathbb{C}$ on which $\Omega_\alpha(z)$ is defined, where we set $I_0=I_{n+1}=\emptyset$. 
By adding the point $z=\infty$ to each $\Sigma_\alpha$, we obtain two disks from $\Sigma_0$ and $\Sigma_{n}$, and $n-1$ cylinders from the other $\Sigma_\alpha$. 
From these pieces, we obtain $\mathbb{CP}^1$ by gluing them along the cuts $I_\alpha$. 
This procedure is depicted in Figure \ref{gluing to CP1}. 
The equations (\ref{gluing}) then imply that $\Omega_\alpha(z)$ consistently define a function $\Omega(s)$ on $\mathbb{CP}^1$, where $s$ is a coordinate of $\mathbb{CP}^1$. 

Suppose that $\Omega(s)$ had been determined. 
Then, $\Omega_\alpha(z)$ can be recovered by composing $\Omega(s)$ with the map $s_\alpha:\Sigma_\alpha\to\mathbb{CP}^1$ which was used for the gluing. 
To obtain $zv_a'(z)$ from $\Omega_\alpha(z)$ is then a trivial task. 
A difficulty in this strategy for obtaining $v'_a(z)$ is that the maps $s_\alpha$ seem to be quite complicated. 
Indeed, each of them would be given by a multi-valued function satisfying a polynomial equations of degree $n+1$ with function coefficients. 

Fortunately, it turns out that the inverse of $s_\alpha$ can be described quite explicitly. 
Let $U_\alpha\subset \mathbb{CP}^1$ be the image of $\Sigma_\alpha\cup\{\infty\}$ by the map $s_\alpha$. 
The inverse of $s_\alpha$ defines a function $z|_{U_\alpha}(s)$ on $U_\alpha$. 
This function has a simple pole at $s=s_\alpha(\infty)$. 
Since the closures $\overline{U_\alpha}$ of $U_\alpha$ cover $\mathbb{CP}^1$, we can define a meromorphic function $z(s)$ on $\mathbb{CP}^1$ from $z|_{U_\alpha}(s)$. 
Since any meromorphic function on $\mathbb{CP}^1$ is a rational function, we conclude that $z(s)$ is of the form 
\begin{equation}
z(s)\ =\ D\prod_{\alpha=0}^n\frac{s-\eta_\alpha}{s-\xi_\alpha}. 
   \label{z(s) general}
\end{equation}

The analytic structure of $\Omega(s)$ is determined from those of $v_a(z)$ as follows. 
Recall that $v_a(z)$ are holomorphic except at their branch points. 
This implies that $\Omega(s)$ is holomorphic on $\mathbb{CP}^1$ except at $2n$ points corresponding to the branch points of $v_a(z)$. 
Let $z=p$ be one of the branch point. 
At this point, $v_a(z)$ is expected to behave as 
\begin{equation}
v_a(z)\ \sim\ v_a(p)+c\sqrt{z-p}. 
\end{equation}
Then, the derivative behaves as 
\begin{equation}
zv_a'(z)\ \sim\ \frac{cp}{2\sqrt{z-p}}. 
   \label{origin of pole}
\end{equation}
The same behavior is shared by $\Omega_a(z)$. 

Let $s=\sigma$ be a point on $\mathbb{CP}^1$ such that $z(\sigma)=p$ is satisfied. 
At this point, $z(s)$ should behave as 
\begin{equation}
z(s)\ \sim\ p+c'(s-\sigma)^2
\end{equation}
since $z=p$ is a branch point of $v_a(z)$. 
This implies that the pull-back by $z(s)$ of the functions $\Omega_\alpha(z)$ behaving like (\ref{origin of pole}) has a simple pole at $s=\sigma$. 

We have found that $\Omega(s)$ is a meromorphic function on $\mathbb{CP}^1$ with $2n$ simple poles. 
This is a rational function of the form 
\begin{equation}
\Omega(s)\ =\ A+\sum_{a=1}^n\left[ \frac{B_a}{s-\sigma_a}+\frac{C_a}{s-\tau_a} \right],  
   \label{Omega general}
\end{equation}
where $z(\sigma_a)$ and $z(\tau_a)$ are the branch points of $v_a(z)$. 

The observables discussed in section \ref{preliminaries} can be given directly in terms of $\Omega(s)$ and $z(s)$ defined above. 
For example, the vevs $W_a$ of the Wilson loops are given as 
\begin{equation}
t_{\alpha+1}W_{\alpha+1}-t_\alpha W_\alpha\ =\ -\frac12D(\xi_\alpha-\eta_\alpha)\Omega'(\xi_\alpha)\prod_{\beta\ne \alpha}\frac{\xi_\alpha-\eta_\beta}{\xi_\alpha-\xi_\beta}, 
   \label{WL intermediate}
\end{equation}
and the 't~Hooft couplings $t_a$ are given as 
\begin{equation}
t_{\alpha+1}-t_\alpha\ =\ \frac12\int_{\eta_\alpha}^{\xi_\alpha}ds\,\frac{z'(s)}{z(s)}\left( \Omega(s)-\Omega_\alpha \right), \hspace{1cm} \Omega_\alpha\ :=\ \Omega(\xi_\alpha). 
   \label{t Hooft intermediate}
\end{equation}
Note that we set $t_0=t_{n+1}=W_0=W_{n+1}=0$, so that $t_a$ and $W_a$ are completely determined by (\ref{WL intermediate}) and (\ref{t Hooft intermediate}). 
By eliminating some of the parameters, we obtain $W_a$ as functions of $t_a$. 
The integrands for the 't~Hooft couplings are in fact rational functions, and therefore the integration can be done readily. 
The results are given in terms of the parameters in $\Omega(s)$ and $z(s)$ which are to be determined. 

The free energy $F$ can be also rewritten in terms of $\Omega(s)$ and $z(s)$. 
Note that in any case we do not need to deal with the inverse of $z(s)$ for the calculations of the observables.

\vspace{5mm}

\subsection{Fundamental 1-form}

\vspace{5mm}

We need to determine $\Omega(s)$ and $z(s)$. 
It turns out that there are constraints on them deduced from the properties of $v_a(z)$, by which enough number of the parameters in $\Omega(s)$ and $z(s)$ can be determined. 
This is shown in Appendix \ref{constraints}. 
The undetermined parameters are related to $t_a$, as explained at the end of section \ref{preliminaries}. 
Therefore, we conclude that we can solve the equations (\ref{saddle point derivative 2}) completely, in principle. 
In practice, the values of the parameters are given as solutions of a set of algebraic equations. 
The existence of the solution can be shown, but it is almost impossible to understand the analytic behaviors of them as functions of $t_a$ for general values of $n$. 

Fortunately, there is a shortcut. 
It turns out that all the information we need to know is encoded in the following 1-form 
\begin{equation}
\Lambda\ :=\ \frac{z'(s)}{z(s)}\Omega(s)ds. 
   \label{fundamental 1-form def}
\end{equation}
It is curious to notice that $\Lambda$ is obtained from the pull-back of a 1-form $dv_{\alpha+1}-dv_\alpha$ on $\Sigma_\alpha$ to $\mathbb{CP}^1$ by $z(s)$ as 
\begin{equation}
z^*(dv_{\alpha+1}-dv_\alpha)\ =\ \Lambda-\Omega_\alpha d\log z(s). 
\end{equation}
The advantage of considering $\Lambda$ instead of $\Omega(s)$ and $z(s)$ separately is that $\Lambda$ has a very simple expression given below. 

Recall that $\Omega(s)$ has simple poles at $s=\sigma_a,\tau_a$. 
Since $z(\sigma_a)$ and $z(\tau_a)$ correspond to the branch points of $v_a(z)$, the values of $\sigma_a$ and $\tau_a$ are determined as the solutions of $z'(s)=0$. 
Therefore, the poles of $\Omega(s)$ are canceled by the zeros of $z'(s)$ in $\Lambda$. 
There are other poles coming from 
\begin{equation}
\frac{z'(s)}{z(s)}\ =\ \sum_{\alpha=0}^n\left( \frac1{s-\eta_\alpha}-\frac1{s-\xi_\alpha} \right). 
\end{equation}
Therefore, $\Lambda$ is given simply as the sum of these poles with appropriate residues. 
We obtain 
\begin{equation}
\Lambda\ =\ \sum_{\alpha=0}^n\left( \frac{\Omega_\alpha}{s-\eta_\alpha}-\frac{\Omega_\alpha}{s-\xi_\alpha} \right)ds, 
   \label{fundamental 1-form}
\end{equation}
where we have used $\Omega(\eta_\alpha)=\Omega_\alpha$, which is shown in Appendix \ref{constraints}. 
The residues $\Omega_\alpha$ are given as 
\begin{equation}
\Omega_\alpha\ =\ c_\alpha-c_{\alpha+1}, 
\end{equation}
and $c_\alpha$ are given in terms of $\kappa_a$ by (\ref{def of c_a}). 

Now, the 't~Hooft couplings $t_a$ can be obtained very explicitly as 
\begin{equation}
t_{\alpha+1}-t_\alpha\ =\ \frac12\sum_{\beta\ne\alpha}(\Omega_\beta-\Omega_\alpha)\log r_{\alpha\beta}, 
   \label{formula t Hooft}
\end{equation}
where $r_{\alpha\beta}$ are the cross ratios 
\begin{equation}
r_{\alpha\beta}\ :=\ \frac{\xi_\alpha-\eta_\beta}{\eta_\alpha-\eta_\beta}\frac{\eta_\alpha-\xi_\beta}{\xi_\alpha-\xi_\beta}, 
\end{equation}
and the coefficients $\Omega_\beta-\Omega_\alpha$ can be given simply as 
\begin{equation}
\Omega_\beta-\Omega_\alpha\ =\ 2\sum_{\gamma=\alpha+1}^\beta\kappa_\gamma
\end{equation}
for $\beta>\alpha$. 

We need to know $\Omega'(\xi_\alpha)$ which is necessary in (\ref{WL intermediate}) for the vevs $W_a$. 
It turns out that these quantities can be also obtained from $\Lambda$. 
By expanding (\ref{fundamental 1-form def}) at $s=\xi_\alpha$, we find 
\begin{equation}
\Lambda\ =\ \left[ -\frac{\Omega_\alpha}{s-\xi_\alpha}-\Omega'(\xi_\alpha)+\frac{\Omega_\alpha}{\xi_\alpha-\eta_\alpha}+\sum_{\beta\ne\alpha}\left( \frac{\Omega_\alpha}{\xi_\alpha-\eta_\beta}-\frac{\Omega_\alpha}{\xi_\alpha-\xi_\beta} \right)+{\cal O}(s-\xi_\alpha) \right]ds, 
\end{equation}
On the other hand, by expanding (\ref{fundamental 1-form}), we find 
\begin{equation}
\Lambda\ =\ \left[ -\frac{\Omega_\alpha}{s-\xi_\alpha}+\frac{\Omega_\alpha}{\xi_\alpha-\eta_\alpha}+\sum_{\beta\ne\alpha}\left( \frac{\Omega_\beta}{\xi_\alpha-\eta_\beta}-\frac{\Omega_\beta}{\xi_\alpha-\xi_\beta} \right)+{\cal O}(s-\xi_\alpha) \right]ds, 
\end{equation}
Comparing these two expressions, we obtain 
\begin{equation}
\Omega'(\xi_\alpha)\ =\ \sum_{\beta\ne\alpha}\left( \frac{\Omega_\alpha-\Omega_\beta}{\xi_\alpha-\eta_\beta}-\frac{\Omega_\alpha-\Omega_\beta}{\xi_\alpha-\xi_\beta} \right). 
\end{equation}
Therefore, the vevs $W_a$ can be given as 
\begin{equation}
t_{\alpha+1}W_{\alpha+1}-t_\alpha W_\alpha\ =\ -\frac12D(\xi_\alpha-\eta_\alpha)\sum_{\beta\ne\alpha}\left( \frac{\Omega_\alpha-\Omega_\beta}{\xi_\alpha-\eta_\beta}-\frac{\Omega_\alpha-\Omega_\beta}{\xi_\alpha-\xi_\beta} \right)\prod_{\gamma\ne\alpha}\frac{\xi_\alpha-\eta_\gamma}{\xi_\alpha-\xi_\gamma}. 
   \label{formula Wilson}
\end{equation}

The free energy $F$ can be also written in terms of $\Lambda$. 
As a result, we obtain $F$ as a sum of integrals whose integrands are combinations of rational functions and logarithms. 

\vspace{1cm}

\section{Examples}   \label{examples}

\vspace{5mm}

In this section, we examine the implications of the formulas (\ref{formula t Hooft})(\ref{formula Wilson}) obtained in the last section. 
We will consider the cases $n=1$ and $n=2$. 
In the following, we employ 
\begin{equation}
z(s)\ =\ -\prod_{\alpha=0}^n\frac{\xi_\alpha s-1}{s-\xi_\alpha}. 
   \label{z(s) simple}
\end{equation}
Note that this corresponds to taking $\eta_\alpha=\xi_\alpha^{-1}$. 
In Appendix \ref{constraints}, we show that this choice gives a rather simple solution to the equations (\ref{saddle point derivative 2}). 
Note that the choice of the coordinate on $\mathbb{CP}^1$ is not physically meaningful, and they can be changed by an ${\rm SL}(2,\mathbb{C})$ transformation. 
After fixing $\eta_\alpha=\xi_\alpha^{-1}$ as above, there is still a freedom to make an ${\rm SL}(2,\mathbb{C})$ transformation which brings one of $\xi_\alpha$ to infinity. 
In Appendix \ref{constraints}, we choose $\xi_0=\infty$. 

\vspace{5mm}

\subsection{Pure Chern-Simons theory}

\vspace{5mm}

This theory corresponds to $n=1$. 
In this case, we can choose $k=k_1$, and then we have $\kappa_1=1$. 
For the simplicity of notation, we denote $t=t_1$. 
The equations (\ref{saddle point derivative 2}) become 
\begin{equation}
2\ =\ xv'(x_+)+xv'(x_-), \hspace{1cm} x\in I. 
   \label{saddle point pureCS}
\end{equation}

The formula (\ref{formula t Hooft}) implies 
\begin{equation}
t\ =\ \log\frac{\xi_0-\eta_1}{\eta_0-\eta_1}\frac{\eta_0-\xi_1}{\xi_0-\xi_1}, \hspace{1cm} \eta_\alpha\ =\ \xi_\alpha^{-1}. 
\end{equation}
As explained above, we can choose $\xi_0=\infty$. 
We denote $\xi=\xi_1$. 
Then, we obtain 
\begin{equation}
t\ =\ \log\xi^2. 
\end{equation}
The weak coupling limit corresponds to $\xi^2=1$. 
Since we defined that $t=2\pi iN_1/k_1$ is pure imaginary for the physical parameter region, the corresponding $\xi$ is a pure phase. 
The 't~Hooft coupling $t$ grows as $\xi$ rotates around the unit circle in $\mathbb{C}$. 

In fact, it is not possible to decide whether $\xi=+1$ or $\xi=-1$ should correspond to the weak coupling limit, based only on the results on pure Chern-Simons theory. 
In the next section, we will find that $\xi=-1$ is preferable by examining theories with flavors added. 

The vev $W=W_1$ of the Wilson loop is given as 
\begin{equation}
W\ =\ \frac{\xi^2-1}t. 
\end{equation}
By rewriting this as 
\begin{equation}
W\ =\ e^{\frac12t}\frac{e^{\frac12t}-e^{-\frac12t}}t, 
\end{equation}
we find that this reproduces the planar limit of the exact result obtained in \cite{Witten:1988hf} with a framing factor $e^{\frac12t}$ \cite{Kapustin:2009kz}. 
This approaches $1$ in the weak coupling limit, as it should be. 

It is curious to observe that $W$ vanishes at $t=2\pi i$, or in other words, $N=k$. 
This is known to be the boundary of the region in the parameter space in which the supersymmetry is broken \cite{Kitao:1998mf,Bergman:1999na,Ohta:1999iv}. 
In the next subsection, we will show that a similar phenomenon is observed also for $n=2$. 
However, this is not a general property of the vevs of the Wilson loop. 
We will see a counter-example in section \ref{flavors}

\vspace{5mm}

\subsection{Theories for $n=2$}

\vspace{5mm}

The 't~Hooft couplings are given as 
\begin{eqnarray}
t_1 
&=& \kappa_1\log r_{01}+(\kappa_1+\kappa_2)\log r_{02}, \\
t_2 
&=& (\kappa_1+\kappa_2)\log r_{20}+\kappa_2\log r_{21}. 
\end{eqnarray}
The analysis for generic parameters looks to be involved. 
For simplicity, we consider the case $k_1+k_2=0$, for which the supersymmetry is enhanced to ${\cal N}=4$ \cite{Gaiotto:2008sd}. 
For this case, we choose $k=k_1$ so that we have $\kappa_1=-\kappa_2=1$. 

We find that the choice $\xi_1=\infty$ is the most convenient. 
For this choice, we obtain 
\begin{equation}
t_1\ =\ \log(\xi_0)^2, \hspace{1cm} t_2\ =\ -\log(\xi_2)^2. 
\end{equation}
As explained in Appendix \ref{constraints}, the parameters $\xi_0$ and $\xi_2$ are unconstrained.  The physical parameter space then corresponds to $S^1\times S^1$ given by $|\xi_0|=|\xi_2|=1$. 

The vevs of the Wilson loops are given as 
\begin{eqnarray}
W_1 
&=& \frac{1}{t_1}\frac{(\xi_0^2-1)(\xi_0\xi_2-1)}{\xi_0-\xi_2}, \\
W_2 
&=& -\frac{1}{t_2}\frac{(\xi_2^2-1)(\xi_0\xi_2-1)}{\xi_0-\xi_2}. 
\end{eqnarray}
To express them in terms of $t_1$ and $t_2$, we need to fix the sign ambiguity in the relation between $\xi_0$ ($\xi_2$) and $t_1$ ($t_2$), respectively. 
This can be done by checking whether both $W_1$ and $W_2$ approach 1 as $t_1$ and $t_2$ go to zero. 
We find that the correct choice of the signs is 
\begin{equation}
\xi_0\ =\ -e^{\frac12t_1}, \hspace{1cm} \xi_2\ =\ e^{-\frac1{2}t_2}. 
\end{equation}

Then, the vevs are given as 
\begin{eqnarray}
W_1 
&=& \frac{e^{t_1}-1}{t_1}\frac{e^{\frac12(t_1-t_2)}+1}{e^{\frac12t_1}+e^{-\frac12t_2}}, \\
W_2 
&=& -\frac{e^{-t_2}-1}{t_2}\frac{e^{\frac12(t_1-t_2)}+1}{e^{\frac12t_1}+e^{-\frac12t_2}}. 
\end{eqnarray}
The perturbative expansions of these expressions are given as 
\begin{eqnarray}
W_1 
&=& 1+\frac{t_1}2+\frac{t_1^2}6-\frac{t_1t_2}8+{\cal O}(t^3), \\
W_2 
&=& 1-\frac{t_2}2+\frac{t_2^2}6-\frac{t_1t_2}8+{\cal O}(t^3). 
\end{eqnarray}
They reproduce the perturbative results found in \cite{Suyama:2016nap} which were obtained directly from the localization formula (\ref{localized Wilson loop}). 

\vspace{5mm}

The vevs $W_1$ and $W_2$ have interesting non-perturbative behaviors. 
Since the structures of $W_1$ and $W_2$ are almost identical, we focus on $W_1$. 

$W_1$ has a factor which also appeared in $W$ for pure Chern-Simons theory. 
Therefore, $W_1$ vanishes whenever $N_1=k_1$. 
There is another factor in the numerator which vanishes when $\frac12(t_1-t_2)=\pi i$ mod $2\pi i$, or in other words, 
\begin{equation}
N_1-N_2\ \in\ (2\mathbb{Z}+1)k. 
   \label{vanishing condition}
\end{equation}
It can be shown by the s-rule \cite{Hanany:1996ie} that the supersymmetry of the theory is broken when $|N_1-N_2|=k$. 
This condition for the supersymmetry breaking coincides with the vanishing condition (\ref{vanishing condition}) of $W_1$. 

Curiously, the denominator of $W_1$ vanishes when $\frac12(t_1+t_2)=\pi i$ mod $2\pi i$, or 
\begin{equation}
N_1+N_2\ \in\ (2\mathbb{Z}+1)k. 
\end{equation}
Note that one of them, $N_1+N_2=k$, was discussed in \cite{Nosaka:2017ohr}\cite{Nosaka:2018eip} as the boundary between the good theories and bad theories \cite{Gaiotto:2008ak}. 
It would be interesting to investigate this singularity in more detail. 
It would be also interesting to investigate the parameter region beyond the singularity by the analytic continuation of the parameters. 
This is indeed possible since the singularity is a pole for $t_1+t_2$ which can be avoided by taking complex  values of $t_1$ and $t_2$. 

\vspace{1cm}

\section{Adding flavors} \label{flavors}

\vspace{5mm}

It is rather straightforward to analyze theories which are obtained by adding an arbitrary number of fundamental hypermultiplets to the theories considered so far. 
If we add a hypermultiplet belonging to the fundamental representation of ${\rm U}(N_a)$, the localization formula for the partition function (\ref{localized partition function}) includes the factor 
\begin{equation}
\prod_{i=1}^{N_a}\left( \cosh\frac{u^a_i}2 \right)^{-1}. 
\end{equation}
Let $n_a$ be the number of such fundamental hypermultiplets. 
Then, the resolvents $v_a(z)$ defined by (\ref{resolvents}) satisfy 
\begin{equation}
2\kappa_a+2\epsilon_a\nu_a\frac{x}{(x+\epsilon_a)^2}\ =\ xv_a'(x_+)+xv_a'(x_-)-xv_{a-1}'(x)-xv_{a+1}'(x), 
\end{equation}
where we have defined $\epsilon_a$ and $\nu_a$ as 
\begin{equation}
\epsilon_a\ :=\ -(-1)^a, \hspace{1cm} \nu_a\ :=\ \frac{2\pi in_a}{k}. 
\end{equation}

To solve these equations, we decompose $zv_a'(z)$ as 
\begin{equation}
zv_a'(z)\ =\ c_a+d_a(z)+\omega_a(z), 
\end{equation}
where $c_a$ are defined by (\ref{def of c_a}) as before, and $d_a(z)$ are rational functions satisfying 
\begin{equation}
2\epsilon_a\nu_a\frac{z}{(z+\epsilon_a)^2}\ =\ 2d_a(z)-d_{a-1}(z)-d_{a+1}(z). 
\end{equation}
The solution of these equations has of the form 
\begin{equation}
d_a(z)\ =\ \frac{d_{a,-2}}{(z-1)^2}+\frac{d_{a,-1}}{z-1}+\frac{\tilde{d}_{a,-2}}{(z+1)^2}+\frac{\tilde{d}_{a,-1}}{z+1}+d_{a,0}. 
\end{equation}
Since $zv_a'(z)$ are expected to be holomorphic at $z=\pm1$, $\omega_a(z)$ are required to have poles of the form 
\begin{equation}
\omega_a(z)\ =\ -\frac{d_{a,-2}}{(z-1)^2}-\frac{d_{a,-1}}{z-1}-\frac{\tilde{d}_{a,-2}}{(z+1)^2}-\frac{\tilde{d}_{a,-1}}{z+1}+\mbox{(holomorphic)}. 
\end{equation}

The 't~Hooft couplings are given as 
\begin{equation}
t_a\ =\ \frac12\int_0^\infty\frac{dz}{z}\,(c_a+\omega_a(z))+t^{(\nu)}_a, 
\end{equation}
where $t^{(\nu)}_a$ are the solutions of the equations 
\begin{equation}
\nu_a\ =\ 2t_a^{(\nu)}-t_{a-1}^{(\nu)}-t_{a+1}^{(\nu)}. 
\end{equation}
The vevs $W_a$ of the Wilson loops are given as 
\begin{equation}
W_a\ =\ -\frac{\epsilon_a}{2t_a}\lim_{z\to\infty}z\cdot (c_a+\omega_a(z))-\frac1{2t_a}w_a, 
\end{equation}
where $w_a$ satisfy 
\begin{equation}
2\nu_a\ =\ 2w_a+w_{a-1}+w_{a+1}. 
\end{equation}

\vspace{5mm}

\subsection{$n=1$}

\vspace{5mm}

Let us illustrate the above procedure for the case $n=1$, that is, for ${\cal N}=3$ ${\rm U}(N)$ Chern-Simons theory coupled to $n_1$ fundamental hypermultiplets. 
The resolvent $v(z)$ satisfy 
\begin{equation}
2+2\nu\frac x{(x+1)^2}\ =\ xv'(x_+)+xv'(x_-), \hspace{1cm} x\in I_1, 
\end{equation}
where $\nu=\nu_1$. 
We decompose $zv'(z)$ into $c+d(z)+\omega(z)$ where 
\begin{equation}
d(z)\ =\ \nu\frac z{(z+1)^2}\ =\ \frac{-\nu}{(z+1)^2}+\frac{\nu}{z+1}. 
\end{equation}
The part $c+\omega(z)$ satisfies (\ref{saddle point pureCS}) for pure Chern-Simons theory. 
Therefore, the determination of $\omega(z)$ is almost the same as before, except for the requirement that $\omega(z)$ must have poles at $z=-1$ of the form 
\begin{equation}
\omega(z)\ =\ \frac{\nu}{(z+1)^2}+\frac{-\nu}{z+1}+\mbox{(holomorphic)}. 
\end{equation}
We use the same formula (\ref{z(s) simple}) for $z(s)$. 
Recall that the necessary object for determining the observables is the fundamental 1-form $\Lambda$, not $\Omega(s)$ itself. 
We find that $\Lambda$ is given as 
\begin{eqnarray}
\Lambda 
&=& \left[ -\frac1{s-\eta_0}+\frac1{s-\xi_0}+\frac1{s-\eta_1}-\frac1{s-\xi_1} \right. \nonumber \\ [2mm] 
& & \left. +\frac{(\xi_0-1)(\xi_1-1)}{2(\xi_0\xi_1-1)}\frac{\mu}{(s-1)^2}+\frac{(\xi_0+1)(\xi_1+1)}{2(\xi_0\xi_1-1)}\frac{\mu}{(s+1)^2} \right]ds, 
\end{eqnarray}
where the second line comes from poles of $\omega(z)$ at $z=1$. 
As for pure Chern-Simons theory, we choose $\xi_0=\infty$. 
Then, all the observables are functions of $\xi=\xi_1$. 
The 't~Hooft coupling $t$ is given as 
\begin{equation}
t\ =\ \log\xi^2+\frac{\nu}{2}\left( 1+\frac1\xi \right), 
   \label{t Hooft-flavor}
\end{equation}
which is the same as the expression obtained in \cite{Suyama:2014sxa}. 
The weak coupling limit corresponds to the limit $\xi\to-1$. 
The vev $W$ of the Wilson loop is given as 
\begin{equation}
W\ =\ \frac1t\left( \xi^2-1-\frac\nu2(1+\xi) \right). 
   \label{WL flavor}
\end{equation}
To write $W$ explicitly as a function of $t$, we need to solve (\ref{t Hooft-flavor}) for $\xi$. 
Since this equation is complicated, we assume that $\nu$ is small, and solve it perturbatively. 
As a result, we obtain 
\begin{equation}
\xi\ =\ -\exp\left[ \frac12\left( t-\frac12(1-e^{-\frac12t})\nu+\frac18e^{-\frac12t}(1-e^{-\frac12t})\nu^2 \right) \right]+{\cal O}(\nu^3). 
\end{equation}
Substituting this into (\ref{WL flavor}), we obtain 
\begin{equation}
W\ =\ \frac{e^t-1}{t}-\frac12\frac{(e^{\frac12t}-1)^2}t\nu+\frac18\frac{(e^{\frac12t}-1)^2}t\nu^2+{\cal O}(\nu^3). 
\end{equation}
We find that $W$ approaches $1$ in the limit $t\to0$, as expected. 

We have observed so far that the vevs of the Wilson loops vanish for values of $t_a$ corresponding to the supersymmetry breaking. 
For the theory under consideration, the condition for the supersymmetry breaking is $N>k+m$ \cite{Suyama:2014sxa}. 
Then, one might expect that $W$ above vanishes when $t=2\pi i+\nu$. 
However, we find 
\begin{equation}
W\Big|_{t=2\pi i+\nu}\ =\ -\nu+{\cal O}(\nu^3). 
\end{equation}
This indicates that the condition for the vanishing of Wilson loop vevs does not always coincide with the condition for the supersymmetry breaking. 
Note that $W$ vanishes $\xi=1+\frac\nu2$ instead. 

\vspace{1cm}

\section{Discussion}   \label{discuss}

\vspace{5mm}

We have investigated the linear Chern-Simons-matter theories in the planar limit, based on the localization formula (\ref{localized partition function}) of the partition function. 
We have found that all the information necessary for obtaining the free energy $F$ and the vevs $W_a$ of Wilson loops in the planar limit is encoded into a single 1-form $\Lambda$ on an auxiliary $\mathbb{CP}^1$ which we called the fundamental 1-form of the theory. 
The form of $\Lambda$ was determined explicitly. 
As a result, we obtained the formulas for $W_a$ as functions of the 't~Hooft couplings. 
The analytic behavior of $W_a$ was investigated in detail for $n=1$ and $n=2$. 
For the latter case, $W_a$ may vanish or diverge for finite values of $t_a$. 
This seems to be related to some non-perturbative physics, for example, the one discussed in \cite{Nosaka:2017ohr}\cite{Nosaka:2018eip}. 
For $F$, we obtained an integral formula in which the integrand is given by an elementary function. 

It would be interesting to investigate the linear theories for $n>2$ in more detail. 
For those theories, the formulas for the observables were already obtained in this paper. 
However, even though they are given in terms of elementary functions, their analysis does not look straightforward. 
For example, the physical parameter subspace in $\mathbb{CP}^{2n-1}$ would be quite involved for general $n$, although for $n=1$ and $n=2$ it is simply $S^1$ and $T^2$, respectively. 
It would be interesting to understand what kinds of zeros and poles for the observables exist in the physical parameter subspace. 
If a non-trivial analytic structure exists, it is important to identify physical properties underlying it. 
In particular, the investigation of the free energy would be important since it contains important information such as the conditions for the supersymmetry breaking \cite{Morita:2011cs}. 

It would be interesting to extend the analysis in this paper to more general Chern-Simons-matter theories in the planar limit. 
The extension to a generalization \cite{Imamura:2008nn}\cite{Jafferis:2008qz} of ABJM theory would be rather straightforward. 
It would be accomplished by replacing the auxiliary $\mathbb{CP}^1$ to $T^2$, and therefore the roles played by rational functions would be taken by elliptic functions. 
The analysis would be almost parallel at least when $\sum_ak_a=0$ holds. 
For a general case, we probably need the logarithm of a theta function as a part of the resolvents. 

There could be further generalization. 
In \cite{Suyama:2016nap}, we studied a family of Chern-Simons-matter theories which is another generalization of the linear theory with $n=1$. 
In the analysis, not functions on $T^2$ but sections of a non-trivial line bundle on $T^2$ played a crucial role. 
The use of such non-trivial line bundles would enable us to analyze more general Chern-Simons-matter theories. 
It is interesting if the structure of the Kac-Moody algebra found in \cite{Suyama:2017rfh} plays some role in the generalizations. 

We have only considered the results in the planar limit. 
It would be interesting to discuss non-planar contributions, as in \cite{Drukker:2010nc} for example. 

Finally, it would be interesting if we could obtain some implications on AdS/CFT correspondence based on our planar analysis on Chern-Simons-matter theories. 

\vspace{1cm}

{\bf \Large Acknowledgements}

\vspace{5mm}

This work was supported in part by Grants-in-Aid for Scientific Research (No.16K05329).

\appendix

\vspace{1cm}

\section{Equations for resolvents}   \label{resolvent equations}

\vspace{5mm}

In this appendix, we derive the equations (\ref{saddle point derivative}) satisfied by the resolvents $v_a(z)$. 

We start with the localization formula \cite{Kapustin:2009kz} for the partition function 
\begin{equation}
Z\ =\ \int du\,\exp\left[ {\frac{i}{4\pi}\sum_{a=1}^nk_a\sum_{i=1}^{N_a}(u^a_i)^2} \right]\frac{{\prod_{a=1}^n\prod_{i<j}^{N_a}\sinh^2\frac{u^a_i-u^a_j}{2}}}{{\prod_{a,b=1}^{n}\prod_{i=1}^{N_a}\prod_{j=1}^{N_{b}}\left( \cosh\frac{u^a_i-u^{b}_j}2 \right)^{\frac12n_{ab}}}}, 
   \label{partition function appendix}
\end{equation}
for an ${\cal N}=3$ Chern-Simons-matter theory whose matters consist of bi-fundamental fields. 
Our analysis here is not restricted to the linear theories. 
In the following, we allow $n_{ab}$ to be more general non-negative integers, provided that they satisfy (i) $n_{ab}=n_{ba}$, (ii) $n_{aa}=0$, and (iii) the condition (\ref{bipartite condition}) below. 

The saddle point equations for this integral are 
\begin{equation}
\frac{k_a}{2\pi i}u^a_i\ =\ \sum_{j\ne i}^{N_a}\coth\frac{u^a_i-u^a_j}2-\sum_{b\ne a}^n\frac{n_{ab}}2\sum_{j=1}^{N_b}\tanh\frac{u^a_i-u^b_j}2. 
   \label{saddle point appendix}
\end{equation}
We rewrite these equations into more manageable ones. 
Introduce new variables 
\begin{equation}
z_i^a\ :=\ \epsilon_a\exp(u^a_i), \hspace{1cm} \epsilon_a\ :=\ -(-1)^a. 
\end{equation}
In terms of $z_i^a$, (\ref{saddle point appendix}) can be written as 
\begin{equation}
\frac{k_a}{2\pi i}\log(\epsilon_az^a_i)\ =\ \sum_{j\ne i}^{N_a}\frac{z^a_i+z^a_j}{z^a_i-z^a_j}-\sum_{b\ne a}^n\frac{n_{ab}}2\sum_{j=1}^{N_b}\frac{z^a_i+z^b_j}{z^a_i-z^b_j}. 
   \label{saddle point z}
\end{equation}
Here we have assumed that $n_{ab}$ satisfy the following condition 
\begin{equation}
a\ \equiv\ b\mbox{ mod }2 \hspace{1cm} \Rightarrow \hspace{1cm} n_{ab}\ =\ 0. 
   \label{bipartite condition}
\end{equation}
This is satisfied by the linear theories. 
For a theory with $n_{ab}$ satisfying this condition, a diagram corresponding to Figure \ref{diagram} is bipartite. 
Non-bipartite theories were discussed in \cite{Suyama:2013fua}. 

We introduce the resolvents 
\begin{equation}
v_a(z)\ :=\ \frac{t_a}{N_a}\sum_{i=1}^{N_a}\frac{z+z^a_i}{z-z^a_i}, 
   \label{resolvent appendix}
\end{equation}
where $t_a$ are the 't~Hooft couplings defined in (\ref{planar parameters}). 
For finite values of the parameters $k_a$ and $N_a$, these are simple rational functions. 
In the planar limit defined in subsection \ref{planar limit}, it is usually expected that the poles of $v_a(z)$ at $z=z^a_i$ accumulate to form a segment $I_a$ of a curve in $\mathbb{C}$. 
As a result, $v_a(z)$ are expected to become non-trivial functions in the planar limit. 

We assume the following analytic structure of $v_a(z)$ in the planar limit. 
Each $v_a(z)$ is an analytic function on $\mathbb{C}\backslash I_a$. 
On the segment $I_a$, $v_a(z)$ has a branch cut. 
At a branch point, say $z=p_a$, the non-analytic part of $v_a(z)$ behaves as $\sqrt{z-p_a}$, due to the Wigner law. 
In addition, $v_a(z)$ is analytic at infinity which is anticipated from the definition (\ref{resolvent appendix}). 

In terms of the resolvents, the saddle point equations (\ref{saddle point z}) can be written as 
\begin{equation}
2\kappa_a\log(\epsilon_ax)\ =\ v_a(x_+)+v_a(x_-)-\sum_{b\ne a}^nn_{ab}v_b(x), \hspace{1cm} x\in I_a, 
   \label{before derivative}
\end{equation}
where $x_\pm$ are points in the vicinity of $x$ above or below $I_a$, and $\kappa_a$ were defined in (\ref{planar parameters}). 
This rewriting uses the fact that the sum $\frac1{N_a}\sum_{j\ne i}$ becomes the principal value of an integral in the planar limit. 

We have assumed that the segment $I_a$ does not have any intersections with other segments,  and therefore $v_b(x_+)=v_b(x_-)$ hold for $x\in I_a$ with $b\ne a$. 
The resolvents for a more general configuration of the segments are obtained via the analytic continuation of the positions of the branch points. 

The equations (\ref{before derivative}) are not convenient for determining $v_a(z)$ since $\log x$ in the left-hand side requires us to handle the log branch cuts. 
To avoid this difficulty, we take the derivative of the equations \cite{Suyama:2016nap}. 
As a result, we obtain 
\begin{equation}
2\kappa_a\ =\ xv_a'(x_+)+xv_a'(x_-)-\sum_{b\ne a}^nn_{ab}xv_b'(x), \hspace{1cm} x\in I_a. 
   \label{resolvent equations appendix}
\end{equation}
These are the equations analyzed in section \ref{planar solution}. 

\vspace{5mm}

A generalization for adding fundamental hypermultiplets is rather straightforward. 
Let $n_a$ be the number of ${\cal N}=4$ hypermultiplets belonging to the fundamental representation of ${\rm U}(N_a)$. 
The localization formula for such a theory is obtained by inserting the factor 
\begin{equation}
\prod_{a=1}^n\prod_{i=1}^{N_a}\left( \cosh\frac{u^a_i}2 \right)^{-n_a}
\end{equation}
to the integral (\ref{partition function appendix}). 
Formally, this corresponds to adding an extra node $n=0$. 
For the extra node, the level and the rank are $k_0=0$ and $N_0=1$. 
The variable $u^0_1$ is fixed to zero by hand. 
The number $n_a$ is regarded as $n_{0a}$. 

The saddle point equations then become 
\begin{equation}
\frac{k_a}{2\pi i}u^a_i\ =\ \sum_{j\ne i}^{N_a}\coth\frac{u^a_i-u^a_j}2-\sum_{b\ne a}^n\frac{n_{ab}}2\sum_{j=1}^{N_b}\tanh\frac{u^a_i-u^b_j}2-\frac{n_a}2\tanh\frac{u^a_i}2, 
   \label{saddle point flavored}
\end{equation}
where $a=1,2,\cdots,n$. 

The resolvents $v_a(z)$ are defined as in (\ref{resolvent appendix}). 
In terms of them, the saddle point equations (\ref{saddle point flavored}) can be written as 
\begin{equation}
2\kappa_a+2\epsilon_a\mu_a\frac x{(x+\epsilon_a)^2}\ =\ xv_a'(x_+)+xv_a'(x_-)-\sum_{b\ne a}^nn_{ab}xv_b'(x), \hspace{1cm} x\in I_a, 
\end{equation}
where $\nu_a:=2\pi in_a/k$. 
These equations are analyzed in section \ref{flavors}. 

\vspace{1cm}

\section{Planar formulas for observables}   \label{observables}

\vspace{5mm}

The derivatives $v_a'(z)$ contain the information on the observables $F$ and $W_a$. 

The expansion of $v_a(z)$ at infinity is given as 
\begin{equation}
v_a(z)\ =\ t_a\left( 1+\frac2{N_a}\sum_{i=1}^{N_a}z^a_i\cdot z^{-1} \right) +{\cal O}(z^{-2}). 
\end{equation}
Recall that we defined $z^a_i=-(-1)^a\exp(\bar{u}^a_i)$. 
Therefore, $W_a$ can be read off from the coefficient of $z^{-1}$ in $v_a(z)$. 
In terms of $v_a'(z)$, $W_a$ is given as 
\begin{equation}
W_a\ =\ \frac{(-1)^a}{2t_a}\lim_{z\to\infty}z\cdot v_a'(z). 
\end{equation}

The free energy $F$ is given by $S_{\rm eff}[\bar{u}]$, in principle. 
However, the formula for $S_{\rm eff}[\bar{u}]$ in the planar limit is complicated since the double sums in $S_{\rm eff}[\bar{u}]$ become double integrals in the limit. 
There is a relatively simple formula for $F$ using $zv_a'(z)$. 

Recall that $F$ scales as $k^2$ in the planar limit. 
Since $k_a$ and $N_a$ are proportional to $k$, we find that $F$ is a homogeneous function of $k_a$ and $N_a$ of degree 2. 
This implies that $F$ satisfies 
\begin{equation}
F\ =\ \frac12\sum_{a=1}^n\left( k_a\frac{\partial F}{\partial k_a}+N_a\frac{\partial F}{\partial N_a} \right). 
\end{equation}
The $k_a$-derivatives of $F$ can be written as 
\begin{equation}
\frac{\partial F}{\partial k_a}\ =\ \frac1Z\int du\,e^{-S_{\rm eff}[u]}\,\frac{-i}{4\pi}\sum_{i=1}^{N_a}(u^a_i)^2. 
   \label{k-derivative of F}
\end{equation}
Since $v_a(z)$ can be written as 
\begin{equation}
v_a(z)\ =\ 2t_az\cdot\frac1{N_a}\sum_{i=1}^{N_a}\frac1{z-z^a_i}-t_a, 
\end{equation}
the expectation values like (\ref{k-derivative of F}) can be written as 
\begin{equation}
\frac1Z\int du\,e^{-S_{\rm eff}[u]}\,\frac1{N_a}\sum_{i=1}^{N_a}f(e^{u^a_i})\ =\ \int_{C_a}\frac{dz}{2\pi i}\frac{v_a(z)}{2t_az}f(\epsilon_az), 
\end{equation}
where the contour $C_a$ encircles the poles at $z=z^a_i$ but excludes the origin. 
Therefore, $\partial_{k_a}F$ is given as 
\begin{eqnarray}
\frac{\partial F}{\partial k_a} 
&=& -\frac{iN_a}{8\pi t_a}\int_{C_a}\frac{dz}{2\pi i}\frac{v_a(z)}{z}(\log (\epsilon_az))^2 \nonumber \\ [2mm]
&=& \frac{iN_a}{24\pi t_a}\int_{C_a}\frac{dz}{2\pi i}\frac{zv_a'(z)}{z}(\log (\epsilon_az))^3. 
\end{eqnarray}

The $N_a$-derivatives of $F$ can be written as follows: 
\begin{eqnarray}
\frac{\partial F}{\partial N_a} 
&=& -\int_{u_a}^\infty du\,\frac{\partial S_{\rm eff}}{\partial u} \nonumber \\
&=& -\int_{u_a}^\infty du\left[ \frac{k_a}{2\pi i}u-\sum_{i=1}^{N_a}\coth\frac{u-u^a_i}2+\sum_{b\ne a}^n\frac{n_{ab}}2\sum_{i=1}^{N_b}\tanh\frac{u-u^b_i}2 \right], 
\end{eqnarray}
where $u_a$ is an endpoint of $I_a$. 
Note that this integral must be regularized in some manner. 
The quantity $\partial_{N_a}F$ is the change of the free energy $F$ when a single eigenvalue is brought from infinity to an edge of $I_a$. 
The value $u_a$ can be changed to any value on $I_a$ as long as $u^a_i$ are replaced with the saddle point values $\bar{u}^a_i$. 

This can be written in terms of the resolvents as 
\begin{eqnarray}
\frac{\partial F}{\partial N_a} 
&=& -\int_{p_a}^\infty\frac{dz}z\left[ \frac{k_a}{2\pi i}\log (\epsilon_az)-\frac{N_a}{t_a}v_a(z)+\sum_{n\ne a}^n\frac{n_{ab}}2\frac{N_b}{t_b}v_b(z) \right] \nonumber \\
&=& -\frac{k_a}{4\pi i}(\log (\epsilon_az))^2+\log (\epsilon_az)\left( \frac{N_a}{t_a}v_a(z)-\sum_{b\ne a}^n\frac{n_{ab}}2\frac{N_b}{t_b}v_b(z) \right)\Bigg|^\infty_{p_a} \nonumber \\
& & +\int_{p_a}^\infty dz\left[ -\frac{N_a}{t_a}\frac{zv_a'(z)}{z}\log (\epsilon_az)+\sum_{b\ne a}^n\frac{n_{ab}}{2}\frac{N_b}{t_b}\frac{zv_b'(z)}{z}\log (\epsilon_az) \right]. 
   \label{dF/dN}
\end{eqnarray}
Since $\partial_uS_{\rm eff}=0$ is satisfied on the branch cut, most of the surface terms at $z=p_a$ cancels. 
The surface term at $z=\infty$ should be subtracted for the regularization. 
Therefore, the relevant part of the free energy is given by the last integral in (\ref{dF/dN}). 

\vspace{1cm}

\section{Constraints on parameters in $\Omega(s)$ and $z(s)$} \label{constraints}

\vspace{5mm}

We have shown in subsection \ref{pull-back} that the information we need is encoded into two rational functions $\Omega(s)$ and $z(s)$ on $\mathbb{CP}^1$. 
The general form of $\Omega(s)$ is given in (\ref{Omega general}). 
There are $4n+1$ parameters: 
\begin{equation}
A, \hspace{5mm} B_a, \hspace{5mm} C_a, \hspace{5mm} \sigma_a, \hspace{5mm} \tau_a, \hspace{1cm} a\ =\ 1,2,\cdots,n. 
\end{equation}
The general form of $z(s)$ is given in (\ref{z(s) general}). 
There are $2n+3$ parameters: 
\begin{equation}
D, \hspace{5mm} \xi_\alpha, \hspace{5mm} \eta_\alpha, \hspace{1cm} \alpha\ =\ 0,1,2,\cdots,n. 
\end{equation}
We should not fix all the above parameters since $n$ of them will be related to $n$ 't~Hooft couplings $t_a$ by (\ref{t Hooft}). 
Therefore, we need $5n+4$ constraints for fixing the other parameters. 
They are given as follows: 

\begin{itemize}
\item 
The choice of the coordinate $s$ on $\mathbb{CP}^1$ is arbitrary. 
We can make an ${\rm SL}(2,\mathbb{C})$ transformation to fix 3 parameters in $z(s)$. 
\item 
The values of $\sigma_a$ and $\tau_a$, corresponding to the branch points of $v_a(z)$, are determined by 
\begin{equation}
z'(\sigma_a)\ =\ z'(\tau_a)\ =\ 0. 
   \label{branch points}
\end{equation}
Indeed, the equation $z'(s)=0$ has $2n$ solutions since the numerator of $z'(s)$ is a polynomial of degree $2n$. 
The relations between the coefficients of the polynomial and its solutions give $2n$ constraints on $\sigma_a,\tau_a,\xi_\alpha$ and $\eta_\alpha$. 
\item 
The saddle point equations (\ref{saddle point appendix}) are invariant under the sign flip $u^a_i\to-u^a_i$. 
Since the variables $z^a_i$ are given by $\exp(u^a_i)$ up to sign, this invariance implies that the endpoints $z=p_a,q_a$ of the segment $I_a$ must satisfy $p_aq_a=1$. 
These relations give $n$ constraints 
\begin{equation}
z(\sigma_a)z(\tau_a)\ =\ 1. 
   \label{inversion}
\end{equation}
\item 
The resolvents $v_a(z)$ should be holomorphic at $z=0,\infty$. 
This requirement implies that $zv_a'(z)$ vanish at $z=0,\infty$. 
These conditions then require $\Omega(s)$ to satisfy 
\begin{equation}
\Omega(\eta_\alpha)\ =\ \Omega(\xi_\alpha)\ =\ c_\alpha-c_{\alpha+1}. 
   \label{zero and infinity}
\end{equation}
Recall that $c_a$ are determined by (\ref{def of c_a}). 
These give $2n+2$ constraints. 
We will see shortly that only $2n+1$ of them are independent constraints. 
\end{itemize}

In total, we have $3+2n+n+2n+1=5n+4$ constraints, as expected. 

\vspace{5mm}

In the following, we show that the constraints listed above really fix the $5n+4$ parameters. 

First, we show that only $2n+1$ of the constraints (\ref{zero and infinity}) are independent. 
The general form (\ref{Omega general}) of $\Omega(s)$ implies 
\begin{equation}
\sum_{\alpha=0}^n\left( \Omega(\eta_\alpha)-\Omega(\xi_\alpha) \right)\ =\ \sum_{\alpha=0}^n\sum_{a=1}^n\left[ \frac{B_a}{\eta_\alpha-\sigma_a}+\frac{C_a}{\eta_\alpha-\tau_a}-\frac{B_a}{\xi_\alpha-\sigma_a}-\frac{C_a}{\xi_\alpha-\tau_a} \right]. 
\end{equation}
The conditions (\ref{branch points}) can be written as 
\begin{eqnarray}
\sum_{\alpha=0}^n\left[ \frac1{\sigma_a-\eta_\alpha}-\frac1{\sigma_a-\xi_\alpha} \right] &=& 0, \\
\sum_{\alpha=0}^n\left[ \frac1{\tau_a-\eta_\alpha}-\frac1{\tau_a-\xi_\alpha} \right] &=& 0. 
\end{eqnarray}
Using these relations, we find 
\begin{equation}
\sum_{\alpha=0}^\infty\left( \Omega(\eta_\alpha)-\Omega(\xi_\alpha) \right)\ =\ 0. 
\end{equation}
Therefore, at least one equation in (\ref{zero and infinity}) is redundant. 

To show that the remaining equations are independent, we choose $\eta_0=0$ and $\xi_0=\infty$ by using an ${\rm SL}(2,\mathbb{C})$ transformation. 
Then, we obtain $A=\Omega(\eta_0)=\Omega(\xi_0)$. 
The rest of the equations can be written as 
\begin{eqnarray}
\sum_{a=1}^n\left[ \frac{B_a}{\eta_b-\sigma_a}+\frac{C_a}{\eta_b-\tau_a} \right] &=& \Omega(\eta_b)-\Omega(\eta_0), \\
\sum_{a=1}^n\left[ \frac{B_a}{\xi_b-\sigma_a}+\frac{C_a}{\xi_b-\tau_a} \right] &=& \Omega(\xi_b)-\Omega(\xi_0). 
\end{eqnarray}
These equations determine $B_a$ and $C_a$ uniquely if and only if 
\begin{equation}
\frac{\prod_{a<b}^n(\eta_a-\eta_b)(\xi_a-\xi_b)(\sigma_a-\sigma_b)(\tau_a-\tau_b)\prod_{a,b=1}^n(\eta_a-\xi_b)(\sigma_a-\tau_b)}{\prod_{a,b=1}^n(\eta_a-\sigma_b)(\xi_a-\sigma_b)(\eta_a-\tau_b)(\xi_a-\tau_b)}
\end{equation}
is non-vanishing. 
Therefore, generically the $2n+1$ constraints are independent. 

\vspace{5mm}

In fact, a large part of the constraints can be solved explicitly. 
To show this, we choose  
\begin{equation}
z(s)\ =\ s\prod_{a=1}^n\frac{\xi_as-1}{s-\xi_a}. 
   \label{z(s) chosen}
\end{equation}
The freedom to perform the ${\rm SL}(2,\mathbb{C})$ transformations is now fixed. 
This satisfies $z(s^{-1})=z(s)^{-1}$. 
Differentiating this relation, we obtain 
\begin{equation}
-\frac1{s^2}z'(s^{-1})\ =\ -\frac1{z(s)^2}z'(s). 
\end{equation}
This implies that, if $\sigma$ satisfies $z'(\sigma)=0$, then $\tau:=\sigma^{-1}$ also satisfies $z'(\tau)=0$. 
Therefore, by choosing $\tau_a=\sigma_a^{-1}$, the half of (\ref{branch points}) are solved. 
Also, the constraints (\ref{inversion}) are automatically satisfied. 
Then, $\sigma_a$ are determined by solving the equations  
\begin{equation}
\frac1s+\sum_{a=1}^n\left[ \frac1{s-\xi_a^{-1}}-\frac1{s-\xi_a} \right]\ =\ 0. 
\end{equation}

To determine $\Omega(s)$, we choose 
\begin{equation}
B_a\ =\ \sigma_aE_a, \hspace{1cm} C_a\ =\ -\tau_aE_a, 
\end{equation}
using new parameters $E_a$. 
For this choice, $\Omega(s)$ satisfies 
\begin{equation}
\Omega(s^{-1})\ =\ \Omega(s). 
\end{equation}
Since we chose $\eta_a=\xi_a^{-1}$ in (\ref{z(s) chosen}), the half of the conditions (\ref{zero and infinity}) are automatically solved. 
The remaining $n+1$ equations determine $A$ and $E_a$ uniquely. 

\vspace{5mm}

We have shown that the $5n+4$ constraints determine $z(s)$ and $\Omega(s)$. 
All the parameters are determined by $\xi_a$. 
Since there is no further constraint on $\xi_a$, the parameter space is $\mathbb{C}^n$ corresponding to a subspace of $\mathbb{CP}^{2n-1}$ specified by $\kappa_a$ not all of them vanishing. 
Note that there could be a set of discrete choices for the solutions of the constraints. 
We will fix this ambiguity by examining the weak coupling results, as in section \ref{examples}, since they can be also obtained directly from the localized partition function (\ref{partition function appendix}).


\begin{thebibliography}{99}

\bibitem{Hanany:1996ie} 
  A.~Hanany and E.~Witten,
  ``Type IIB superstrings, BPS monopoles, and three-dimensional gauge dynamics,''
  Nucl.\ Phys.\ B {\bf 492}, 152 (1997)
  doi:10.1016/S0550-3213(97)00157-0, 10.1016/S0550-3213(97)80030-2
  [hep-th/9611230].

\bibitem{Kitao:1998mf} 
  T.~Kitao, K.~Ohta and N.~Ohta,
  ``Three-dimensional gauge dynamics from brane configurations with (p,q) - five-brane,''
  Nucl.\ Phys.\ B {\bf 539}, 79 (1999)
  doi:10.1016/S0550-3213(98)00726-3
  [hep-th/9808111].

\bibitem{Bergman:1999na} 
  O.~Bergman, A.~Hanany, A.~Karch and B.~Kol,
  ``Branes and supersymmetry breaking in three-dimensional gauge theories,''
  JHEP {\bf 9910}, 036 (1999)
  doi:10.1088/1126-6708/1999/10/036
  [hep-th/9908075].

\bibitem{Aharony:2008ug} 
  O.~Aharony, O.~Bergman, D.~L.~Jafferis and J.~Maldacena,
  ``N=6 superconformal Chern-Simons-matter theories, M2-branes and their gravity duals,''
  JHEP {\bf 0810}, 091 (2008)
  doi:10.1088/1126-6708/2008/10/091
  [arXiv:0806.1218 [hep-th]].

\bibitem{Imamura:2008nn} 
  Y.~Imamura and K.~Kimura,
  ``On the moduli space of elliptic Maxwell-Chern-Simons theories,''
  Prog.\ Theor.\ Phys.\  {\bf 120}, 509 (2008)
  doi:10.1143/PTP.120.509
  [arXiv:0806.3727 [hep-th]].

\bibitem{Jafferis:2008qz} 
  D.~L.~Jafferis and A.~Tomasiello,
  ``A Simple class of N=3 gauge/gravity duals,''
  JHEP {\bf 0810}, 101 (2008)
  doi:10.1088/1126-6708/2008/10/101
  [arXiv:0808.0864 [hep-th]].

\bibitem{Gaiotto:2008sd} 
  D.~Gaiotto and E.~Witten,
  ``Janus Configurations, Chern-Simons Couplings, And The theta-Angle in N=4 Super Yang-Mills Theory,''
  JHEP {\bf 1006}, 097 (2010)
  doi:10.1007/JHEP06(2010)097
  [arXiv:0804.2907 [hep-th]].

\bibitem{Hosomichi:2008jd} 
  K.~Hosomichi, K.~M.~Lee, S.~Lee, S.~Lee and J.~Park,
  ``N=4 Superconformal Chern-Simons Theories with Hyper and Twisted Hyper Multiplets,''
  JHEP {\bf 0807}, 091 (2008)
  doi:10.1088/1126-6708/2008/07/091
  [arXiv:0805.3662 [hep-th]].

\bibitem{Kapustin:2009kz} 
  A.~Kapustin, B.~Willett and I.~Yaakov,
  ``Exact Results for Wilson Loops in Superconformal Chern-Simons Theories with Matter,''
  JHEP {\bf 1003}, 089 (2010)
  doi:10.1007/JHEP03(2010)089
  [arXiv:0909.4559 [hep-th]].

\bibitem{Nosaka:2017ohr} 
  T.~Nosaka and S.~Yokoyama,
  ``Complete factorization in minimal $ \mathcal{N}=4 $ Chern-Simons-matter theory,''
  JHEP {\bf 1801}, 001 (2018)
  doi:10.1007/JHEP01(2018)001
  [arXiv:1706.07234 [hep-th]].

\bibitem{Nosaka:2018eip} 
  T.~Nosaka and S.~Yokoyama,
  ``Index and duality of minimal $ \mathcal{N} = 4 $ Chern-Simons-matter theories,''
  JHEP {\bf 1806}, 028 (2018)
  doi:10.1007/JHEP06(2018)028
  [arXiv:1804.04639 [hep-th]].

\bibitem{Gaiotto:2007qi} 
  D.~Gaiotto and X.~Yin,
  ``Notes on superconformal Chern-Simons-Matter theories,''
  JHEP {\bf 0708}, 056 (2007)
  doi:10.1088/1126-6708/2007/08/056
  [arXiv:0704.3740 [hep-th]].

\bibitem{Morita:2017oev} 
  T.~Morita and K.~Sugiyama,
  ``Multi-cut solutions in Chern-Simons matrix models,''
  Nucl.\ Phys.\ B {\bf 929}, 1 (2018)
  doi:10.1016/j.nuclphysb.2018.01.028
  [arXiv:1704.08675 [hep-th]].

\bibitem{Suyama:2016nap} 
  T.~Suyama,
  ``Notes on Planar Resolvents of Chern-Simons-matter Matrix Models,''
  JHEP {\bf 1611}, 049 (2016)
  doi:10.1007/JHEP11(2016)049
  [arXiv:1605.09110 [hep-th]].

\bibitem{Suyama:2017rfh} 
  T.~Suyama,
  ``Strong Coupling Limit of A Family of Chern-Simons-matter Theories,''
  JHEP {\bf 1801}, 125 (2018)
  doi:10.1007/JHEP01(2018)125
  [arXiv:1706.08204 [hep-th]].

\bibitem{Witten:1988hf} 
  E.~Witten,
  ``Quantum Field Theory and the Jones Polynomial,''
  Commun.\ Math.\ Phys.\  {\bf 121}, 351 (1989).
  doi:10.1007/BF01217730

\bibitem{Ohta:1999iv} 
  K.~Ohta,
  ``Supersymmetric index and s rule for type IIB branes,''
  JHEP {\bf 9910}, 006 (1999)
  doi:10.1088/1126-6708/1999/10/006
  [hep-th/9908120].

\bibitem{Gaiotto:2008ak} 
  D.~Gaiotto and E.~Witten,
  ``S-Duality of Boundary Conditions In N=4 Super Yang-Mills Theory,''
  Adv.\ Theor.\ Math.\ Phys.\  {\bf 13}, no. 3, 721 (2009)
  doi:10.4310/ATMP.2009.v13.n3.a5
  [arXiv:0807.3720 [hep-th]].

\bibitem{Suyama:2014sxa} 
  T.~Suyama,
  ``Supersymmetry Breaking and Planar Free Energy in Chern-Simons-matter Theories,''
  arXiv:1405.7469 [hep-th].

\bibitem{Morita:2011cs} 
  T.~Morita and V.~Niarchos,
  ``F-theorem, duality and SUSY breaking in one-adjoint Chern-Simons-Matter theories,''
  Nucl.\ Phys.\ B {\bf 858}, 84 (2012)
  doi:10.1016/j.nuclphysb.2012.01.003
  [arXiv:1108.4963 [hep-th]].

\bibitem{Drukker:2010nc} 
  N.~Drukker, M.~Marino and P.~Putrov,
  ``From weak to strong coupling in ABJM theory,''
  Commun.\ Math.\ Phys.\  {\bf 306}, 511 (2011)
  doi:10.1007/s00220-011-1253-6
  [arXiv:1007.3837 [hep-th]].

\bibitem{Suyama:2013fua} 
  T.~Suyama,
  ``A Systematic Study on Matrix Models for Chern-Simons-matter Theories,''
  Nucl.\ Phys.\ B {\bf 874}, 528 (2013)
  doi:10.1016/j.nuclphysb.2013.06.008
  [arXiv:1304.7831 [hep-th]].

\end{thebibliography}
\end{document}